\documentclass[aps,prc,preprint,showpacs,groupedaddress]{revtex4}

\usepackage{graphicx}
\usepackage{amsmath,amssymb,amsfonts}
\usepackage{dcolumn}
\usepackage{bm}
\usepackage[english]{babel}

\newcommand{\ssst}{\scriptscriptstyle}
\newcommand{\sst}{\scriptstyle}
\newcommand{\dst}{\displaystyle}
\newcommand{\abs}[1]{\ensuremath{\vert #1 \vert}}

\begin{document}

\title{Regge-plus-resonance treatment of the \protect\\ $\bm{p(\gamma,K^+)\Sigma^0}$  and $\bm{p(\gamma,K^0)\Sigma^+}$ reactions at forward kaon angles}

\author{T. Corthals}
\email[]{tamara.corthals@ugent.be}
\affiliation{Department of Subatomic and Radiation Physics, \\
  Ghent University, Proeftuinstraat 86, B-9000 Gent, Belgium}
\author{D.G. Ireland}
\affiliation{University of Glasgow, Glasgow G12 8QQ, United Kingdom}
\author{T. Van Cauteren}
\affiliation{Department of Subatomic and Radiation Physics, \\
  Ghent University, Proeftuinstraat 86, B-9000 Gent, Belgium}
\author{J. Ryckebusch}
\affiliation{Department of Subatomic and Radiation Physics, \\
  Ghent University, Proeftuinstraat 86, B-9000 Gent, Belgium}
\date{\today}

\begin{abstract}
An effective-Lagrangian framework for $K\Sigma$ photoproduction from the proton is presented. The proposed model is applicable at forward kaon angles and photon lab energies from threshold up to 16 GeV. 
The high-energy part of the $p(\gamma,K^+)\Sigma^0$ and $p(\gamma,K^0)\Sigma^+$ amplitudes is expressed in terms of Regge-trajectory exchange in the $t$ channel. By supplementing this Regge background with a number of $s$-channel resonances, the model is extended towards the resonance region. The resulting ``Regge-plus-resonance'' (RPR) approach has the advantage that the background contributions involve only a few parameters, which can be largely constrained by the high-energy data. This work compares various implementations of the RPR model, and explores which resonance contributions are required to fit the data presently at hand. It is demonstrated that, through the inclusion of one $K$ and two $K^{\ast}$ trajectories, the RPR framework provides an efficient and unified description of the $K^+\Sigma^0$ and  $K^0\Sigma^+$ photoproduction channels over an extensive energy range. 
\end{abstract}


\maketitle

\section{Introduction}
The goal of attaining a full understanding of the nucleon's internal structure, as reflected for example in its excitation spectrum, is proving to be an elusive one. From the outset, this effort has been thwarted by several complicating factors, the most fundamental being the non-perturbative nature of the strong interaction at hadronic energy scales. Although significant progress has been made in solving QCD on the lattice, the interpretation of dynamical hadronic processes still hinges to a large extent on models containing some
phenomenological ingredients.

In the ongoing search for the link between quark-gluon and hadronic degrees-of-freedom, an impressive amount of effort has been directed towards the study of photo- and electroinduced meson production. Whereas the initial focus of these experiments was mostly on $\pi N$ final states, in recent years the primary interest has shifted to reaction channels like $\gamma^{(\ast)} N \rightarrow \omega N $, $\eta N $, $\pi \pi N$ and $KY$~\cite{LeeSm06}. It is believed that a study of these processes may reveal the existence of some of the ``missing'' resonances, which have been predicted by various constituent-quark models~\cite{CapRob94,Loe01}, but remained unobserved in the $\pi N$ channel. Proof of their existence would constitute a strong confirmation of the validity of the constituent-quark concept.

Associated open-strangeness production reactions are particulary interesting due to the creation of a strange quark-antiquark pair. Over the past years, the $p(\gamma,K)Y$ database~\cite{McNabb03,Zegers03,Glander04} has been supplemented with new high-precision $\gamma p \rightarrow K^+ \Lambda$ and $\gamma p \rightarrow K^+ \Sigma^0$ data from the CLAS~\cite{Brad05}, LEPS~\cite{Sumihama06} and GRAAL~\cite{Lleres06}
collaborations, whereas SAPHIR has provided a new and detailed analysis of the $\gamma p \rightarrow K^0 \Sigma^+$ channel~\cite{Lawall05}. In the light of these data, and with new double polarization~\cite{Schumacher06} and electroproduction~\cite{Carman06} results on the verge of becoming available, we are perhaps nearer than ever to unraveling the $K Y$ production mechanism. 

The treatment of electromagnetic $K Y$ production can be efficiently realized in an effective-field framework, where the particle interactions are modelled by means of effective Lagrangians. A great deal of effort has been devoted to the development of tree-level isobar models, in which the scattering amplitude is constructed from a number of lowest-order Feynman diagrams~\cite{Stijn_sigma,MaBe95,Sara05}. It is obvious that these models have their limitations. They do not explicitly include higher-order mechanisms like channel couplings and final-state interactions. Furthermore, the decay widths commonly introduced to account for the resonances' finite lifetimes, violate the unitarity constraint. In order to resolve some of these problems, one can resort to a coupled-channels analysis, as is done for example in~\cite{Mosel02_pho,MoShklyar05,Diaz05}. However, these analyses also face unresolved challenges, such as accounting for the $\pi \pi N$ channels, which are responsible for about half of the $\gamma N$ total cross section in the higher-mass $N^*$ region.

It can, however, be argued that for many channels a firmly established reaction mechanism is still lacking. Because of the large number of parameters involved, clearing up issues such as the choice of gauge restoration procedure~\cite{US05} or of hadronic form factors in the context of a coupled-channels framework constitutes a gigantic task. For resolving such ambiguities at the level of the individual channels, tree-level models represent a highly valuable asset, precisely because of their relative simplicity. In addition, the extension from photo- to electroproduction is relatively straightforward in a tree-level model, whereas, to our knowledge, no coupled-channels approach to meson electroproduction has as yet been proposed.

In Ref.~\cite{CorthalsL}, we presented a tree-level effective-field model for the $\gamma p \rightarrow K^+ \Lambda$ photoproduction process, valid at forward angles and for photon energies ranging from threshold up to 16 GeV. The model differs from traditional isobar models in its description of the background contribution to the amplitude, which involves the exchange of the $K(494)$ and $K^{\ast}(892)$ Regge trajectories in the $t$ channel. To this Regge background, a number of nucleon resonance diagrams are added. By construction, these resonant contributions vanish at high energies, and a proper high-energy behavior of the amplitude is ensured. An important advantage of this ``Regge-plus-resonance'' (RPR) strategy is that the background coupling constants are heavily constrained by the high-energy $p(\gamma,K^+)\Lambda$ data, leaving the $N^{\ast}$ couplings as the only free parameters in the resonance region.

In this work, the RPR prescription is applied to the $p(\gamma, K)\Sigma$ photoproduction processes. These open a new window onto the hadronic spectrum because $\Delta$ resonances ($\Delta^{\ast}$s) can participate in the reaction
mechanism, unlike $\Lambda$ production. While this implies more resonance candidates to be considered, the number of model parameters in the $K\Sigma$ channels can be kept within bounds by exploiting the isospin relations between the $\gamma p \rightarrow K^+ \Sigma^0$ and  $\gamma p \rightarrow K^0 \Sigma^+$ coupling constants. 

This paper is organized as follows. In Sec.~\ref{sec: RPR} the main ingredients of the RPR model are reviewed. The procedure of $t$-channel reggeization of the high-energy amplitude is discussed in Sec.~\ref{subsec: RPR bg}, whereas Sec.~\ref{subsec: RPR res} elucidates how the constructed model can be extended to the resonance region by the addition of a limited set of $s$-channel resonances. The RPR formalism is then applied to a common description of the two $\gamma p \rightarrow K \Sigma$ processes.  The high-energy Regge amplitudes are constructed in Sec.~\ref{subsec: results bg}, with Sec.~\ref{subsubsec:K+S0} focusing on $\gamma p \rightarrow K^+ \Sigma^0$ and Sec.~\ref{subsubsec:K0S+} on $\gamma p \rightarrow K^0 \Sigma^+$. In Sec.~\ref{subsec: results res} we look for the appropriate $N^{\ast}$ and $\Delta^{\ast}$ resonances with which to supplement the Regge background, and discuss to what extent the presented model succeeds in reproducing the available resonance-region data. Finally, in Sec.~\ref{sec: conclusion} we state our conclusions.

\section{The RPR model}
\label{sec: RPR}

\subsection{$\bm{t}$-channel reggeization}
\label{subsec: RPR bg}

\begin{figure}[t]
\begin{center}
\includegraphics[width=0.45\textwidth]{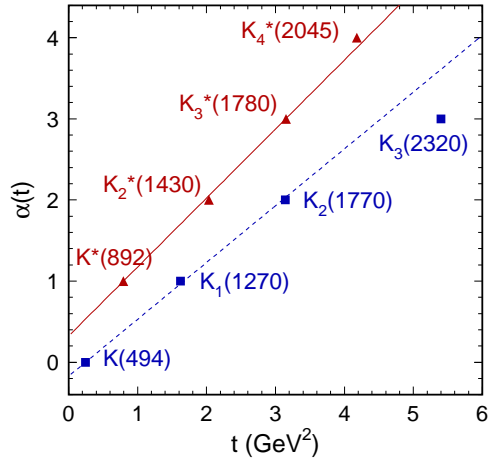}
\caption{(Color online) Chew-Frautschi plots for the $K(494)$ and $K^{\ast}(892)$ trajectories. The meson masses are from the Particle Data Group~\cite{PDG04}.}
\label{fig: kaon-traj}
\end{center}
\end{figure}

Since its initial formulation as an alternative approach to quantum-mechanical potential scattering, Regge theory has been extended far beyond its original scope. Regge's starting-point was to consider the scattering amplitude as a function of a \emph{complex} angular momentum variable~\cite{Regge59}. Interestingly, poles of the scattering amplitude turned out to correspond to resonant states. Regge theory further leads to a natural classification of these resonances into a number of \emph{families}, with identical internal quantum numbers but different spins $J$. Empirically, the hadronic spectrum is observed to exhibit the property that members of such a family, or ``Regge trajectory'', are related by an approximately linear relation $J_i = \alpha(m_i^2)$ between their spins and squared masses. Figure~\ref{fig: kaon-traj} illustrates this point for the $K(494)$ and $K^{\ast}(892)$ trajectories.

The underlying philosophy of the Regge formalism is as follows. In
modelling the reaction amplitude for the $\gamma p \rightarrow K Y$
process at high energies, instead of considering the exchange of a
finite selection of individual particles, the exchange of entire Regge
\emph{trajectories} is taken into account. This exchange can take place in the $t$ channel ($K^{\ast}$ trajectories) or $u$ channel ($Y^{\ast}$ trajectories). As such, Regge theory offers an elegant way to circumvent the controversial issue of modelling high-spin, high-mass particle exchange.
The Regge framework employed here applies to the so-called ``Regge limit'' of extreme forward (in the case of $t$-channel exchange) or backward (for $u$-channel exchange) scattering angles, corresponding to small $\abs{t}$ or $\abs{u}$, respectively. We focus on the forward-angle kinematical region, implying $t$-channel trajectory exchange.

Since the scope of our study extends from parameterizing the $p(\gamma,K)\Sigma$ amplitudes at high energies to unravelling their detailed structure in the resonance region, we have opted to incorporate the Regge formalism into a tree-level effective-Lagrangian model. This approach was pioneered by Guidal and Vanderhaeghen in their treatment of high-energy electromagnetic $\pi$ and $K$ production~\cite{Guidalthes,reg_guidal,reg_guidal_2}. In recent years, several alternative implementations of Regge phenomenology have been proposed, such as the reggeized unitary isobar model~\cite{azna} and quark-gluon strings model~\cite{grish05}. For conciseness, only the basic ingredients of the Regge framework are recalled here. A detailed derivation can be found in Ref.~\cite{CorthalsL} and references therein. 

A linear meson trajectory $\alpha_X(t)$, of which the lightest member (or ``first materialization'') $X$ has a mass $m_X$ and spin $\alpha_{X,0}$, is of the form
\begin{equation}
\alpha_X(t)=\alpha_{X,0} + \alpha'_X \, (t-m_X^2).
\label{eq: reggetraj}
\end{equation}
The amplitude for $t$-channel exchange of the $\alpha_X(t)$ trajectory can be straightforwardly obtained, starting from the standard Feynman amplitude for exchange of its first materialization $X$. The procedure amounts to replacing the denominator of the Feynman propagator with a Regge propagator:
\begin{equation}
\frac{1}{t-m_X^2} \hspace{6pt} \xrightarrow{~\quad} \hspace{8pt} \mathcal{P}^X_{Regge}[s,\alpha_X(t)]\,.
\end{equation}
The Regge amplitude can then be written as
\begin{equation}
\mathcal{M}^X_{Regge}(s,t) = \mathcal{P}^X_{Regge}[s,\alpha_X(t)] ~\times~ \beta_X(s,t) \,,
\label{eq: define_reggeprop}
\end{equation}
with $\beta_X(s,t)$ the residue of the original Feynman amplitude,
to be calculated from the interaction Lagrangians at the $\gamma K X$ and $p X Y$ vertices. 
For spinless external particles, the Regge propagator can be written as~\cite{Donnachie02}:
\begin{equation}
\mathcal{P}^X_{Regge}(s,t) = \frac{ \left(\frac{\dst s}{\dst s_0}\right)^{\alpha_X(t)}}{\sin\bigl(\pi\alpha_X(t)\bigr)} \
\left\{ \begin{array}{c}
1 \\
e^{-i\pi\alpha_X(t)}
\end{array}\right\} \
\frac{\pi\alpha_X'}{\Gamma\bigl(1 + \alpha_X(t)\bigr)}\,,
\label{eq: reggeprop_spinless_degen}
\end{equation}
with the scale factor $s_0$ fixed at $1~\mathrm{GeV}^2$. Eq.~(\ref{eq: reggeprop_spinless_degen}) has poles at nonnegative integer values of $\alpha_X(t)$, hence the interpretation that the
Regge propagator effectively incorporates the exchange of all members
of a trajectory. In the physical plane of the processes under study
($t < 0$), these poles are never reached. 

Expression~(\ref{eq: reggeprop_spinless_degen}) is valid only for
so-called \emph{strongly degenerate} trajectories. In principle, each
Regge trajectory consists of two ``signature parts'', grouping the
trajectory members with positive and negative parity, respectively, and
a separate Regge amplitude should be accorded to each signature
part. Often, though, the two trajectory parts are observed to be
approximately degenerate. This is true for example for the $K(494)$ and
$K^{\ast}(892)$ trajectories, as can be appreciated from
Fig.~\ref{fig: kaon-traj}. In that case, it can often be assumed that the positive- and negative-signature amplitudes have identical \emph{residues}, up to an unknown sign; this is
referred to as \emph{strong} degeneracy. The two can then be added into a single propagator, incorporating the simultanous exchange of both trajectory parts. As is seen from Eq.~(\ref{eq: reggeprop_spinless_degen}), the phase of this propagator can either be constant~(1) or rotating~($e^{-i\pi\alpha(t)}$), depending on the relative sign between the residues of the individual signature parts. For all propagators needed in the treatment of electromagnetic $KY$ production on the proton, the assumption of degenerate trajectories turns out to be a valid one~\cite{Guidalthes, reg_guidal,reg_guidal_2}. 
 
Generalizing Eq.~(\ref{eq: reggeprop_spinless_degen}) to nonscalar particles is a nontrivial task~\cite{Collins77}. We adopt a pragmatic approach, which consists of the following replacement in the spinless-particle propagator of Eq.~(\ref{eq: reggeprop_spinless_degen}): 
\begin{equation}
\alpha_X(t) \longrightarrow \alpha_X(t)-\alpha_{X,0}\,,
\end{equation} 
both in the exponent of $s$ and in the argument of the gamma function. This recipe ensures that the altered propagator has poles at the physical manifestations of the trajectory, i.e. for $\alpha_X(t) \ge \alpha_{X,0}$. The general Regge propagator thus takes the form: 
\begin{equation}
\mathcal{P}^X_{Regge}(s,t) = \frac{ \left(\frac{\dst s}{\dst s_0}\right)^{\alpha_X(t)-\alpha_{X,0}}}{\sin\bigl(\pi\alpha_X(t)\bigr)} \
\left\{ \begin{array}{c}
1 \\
e^{-i\pi\alpha_X(t)}
\end{array}\right\} \
\frac{\pi\alpha_X'}{\Gamma\bigl(1 + \alpha_X(t) -\alpha_{X,0}\bigr )}\,.
\label{eq: reggeprop_spin_degen}
\end{equation}

The above prescription allows one to construct the high-energy
amplitude by selecting the dominant trajectories in the $t$
channel. It turns out that, for fixed $s$,
$\abs{\mathcal{P}^X_{Regge}(s,t=0)}$ increases with decreasing
$\abs{\alpha_X(0)-\alpha_{X,0}} = \alpha'_X m_X^2$. Because all meson
trajectories have approximately the same slope $\alpha'_X$, as a rule
of thumb those with a low-mass first materialization are assumed
to dominate. It should be kept in mind, though, that the coupling
strengths contained in the residues $\beta_X(t)$ [Eq.~(\ref{eq:
    define_reggeprop})] also play a role. In the kaon sector, $K(494)$ and $K^{\ast}(892)$ are by far the lightest states serving as first materializations, hence their importance in modelling the various $\gamma p \rightarrow K Y$ processes.

\subsection{Including resonance dynamics}
\label{subsec: RPR res}

Regge phenomenology is a high-energy tool by construction, because the Regge amplitude built using the propagator of Eq.~(\ref{eq: reggeprop_spin_degen}) is essentially the asymptotic form of the \emph{full} amplitude in the $s \rightarrow \infty$ limit. The experimental meson production cross sections appear to exhibit this ``asymptotic'' Regge behavior for photon energies down to about 4 GeV~\cite{reg_guidal, reg_guidal_2,Sibi03}. As demonstrated in Refs.~\cite{Schumacher06,Guidalthes,Guidal_elec}, however, even with the asymptotic form of the propagators, the gross features of the forward-angle pion and kaon photo- and electroproduction observables in the resonance region are remarkably well reproduced in a pure $t$-channel Regge model. 

The above considerations have prompted us to adopt an identical Regge
description for both the high-energy amplitude and the background
contribution to the resonance-region amplitude. It is evident that a
model consisting exclusively of background diagrams cannot be expected to account for all aspects of
the reaction dynamics. At low energies, the cross sections exhibit
structures which may reflect the presence of individual
resonances. These are incorporated into the RPR framework by
supplementing the reggeized background with a number of resonant
$s$-channel diagrams. For the latter, standard Feynman propagators are
assumed, in which the resonances' finite lifetimes are taken into account through the substitution
\begin{equation}
s - m_{R}^2 \longrightarrow ~ s - m_{R}^2 + im_{R}\,\Gamma_R
\end{equation}
in the propagator denominators, with $m_R$ and $\Gamma_R$ the
mass and width of the propagating state ($R=N^{\ast},\Delta^{\ast}$). 

Further, the condition is imposed that the resonance amplitudes vanish
at high values of $\omega_{lab}$. This is accomplished by including a
Gaussian hadronic form factor at the $K\Lambda R$ vertices:
\begin{equation} 
F_{Gauss}(s) = \exp \left\{- \frac{(s-m^2_{R})^2}{\Lambda_{res}^4}\right\}\,,
\label{eq: gaussff}
\end{equation}
with $\Lambda_{res}$ a universal cutoff mass, to be determined from the
resonance-region data. Our motivation for assuming a Gaussian shape is explained in Ref.~\cite{CorthalsL}. 

The RPR amplitude is shown schematically in Fig.~\ref{fig:
  general_regge_exchange}. It involves $t$-channel exchanges of kaon
trajectories as well as $s$-channel Feynman diagrams corresponding to
individual baryon resonances ($R$). In the high-energy regime ($\omega_{lab} \gtrsim 4$ GeV), all resonant contributions vanish by construction, so that only the Regge part of the amplitude remains. 
 
\begin{figure}[ht]
\begin{center}
\includegraphics[width=0.8\textwidth]{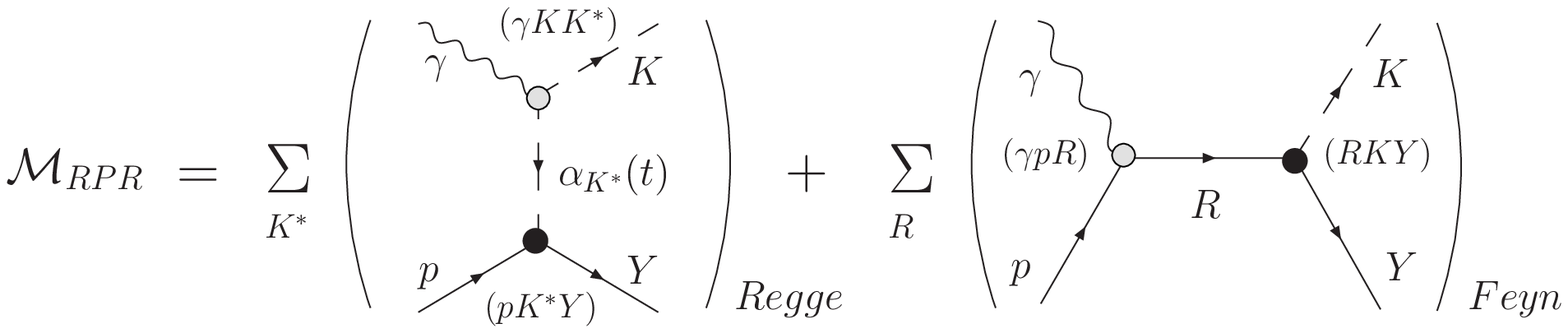}
\caption{General forward-angle RPR amplitude for the $\gamma p \rightarrow K Y$ process.}
\label{fig: general_regge_exchange}
\end{center}
\end{figure}

One issue which may cloud the presented procedure is
double counting. However, because the $\gamma p \rightarrow KY$
processes are largely background-dominated, the few added $s$-channel
terms may be considered as relatively minor corrections, and double
counting is not expected to pose a very serious concern.  

The strong and electromagnetic interaction Lagrangians for coupling to
resonances of various spins are contained in Ref.~\cite{CorthalsL} for
the specific case of $K^+ \Lambda$ photoproduction. Generalizing these Lagrangians to the various $K\Sigma$ channels is fairly straightforward~\cite{corr_lagr}.

\section{$\bm{K\Sigma}$ photoproduction at high energies}
\label{subsec: results bg}

\subsection{The $\bm{\gamma p \rightarrow K^+\Sigma^0}$ process}
\label{subsubsec:K+S0}

We start with the $K^+\Sigma^0$ isospin channel because, to our knowledge, it is the sole one with published high-energy data. As the interaction Lagrangians for the $p(\gamma,K^+)\Lambda$ and $p(\gamma,K^+)\Sigma^0$ processes are essentially identical, the $K^+\Sigma^0$ Regge amplitude can in principle be constructed in complete analogy to the $K^+\Lambda$ one, as presented in our previous work~\cite{CorthalsL}.

Given their low mass, the trajectories corresponding to the $K(494)$ and $K^{\ast}(892)$ mesons are expected to dominate the reaction mechanism. Using the prescription of Eq.~(\ref{eq: reggeprop_spin_degen}), the corresponding Regge propagators read: 
\begin{equation} 
\mathcal{P}^{K(494)}_{Regge}(s,t) = \left(\frac{\dst s}{\dst s_0}\right)^{\alpha_K(t)}
\frac{1}{\sin\bigl(\pi\alpha_K(t)\bigr)}  \hspace{10pt} 
\left\{ \begin{array}{c}
1 \\
e^{-i\pi\alpha_{K}(t)}
\end{array}\right\} \hspace{10pt} 
\frac{\pi \alpha'_K}{\Gamma\bigl(1+\alpha_K(t)\bigr)}\,, \label{eq: reggeprop_K}
\end{equation}
\begin{equation}
\mathcal{P}^{K^{\ast}(892)}_{Regge}(s,t) = \left(\frac{\dst s}{\dst s_0}\right)^{\alpha_{K^{\ast}(892)}(t)-1} 
\frac{1}{\sin\bigl(\pi\alpha_{K^{\ast}(892)}(t)\bigr)} \hspace{10pt}
\left\{ \begin{array}{c}
1 \\
e^{-i\pi\alpha_{K^{\ast}(892)}(t)}
\end{array}\right\} \hspace{10pt}
\frac{\pi \alpha'_{K^{\ast}(892)}}{\Gamma\bigl(\alpha_{K^{\ast}(892)}(t)\bigr)}\,,\label{eq: reggeprop_Kstar}
\end{equation}
with~\cite{Stijnthes} 
\begin{align}
\alpha_K(t) &= 0.70 \ \mathrm{GeV}^{-2} \ (t-m_K^2)\,,\label{eq: Ktraj}\\
\alpha_{K^{\ast}(892)}(t) &= 1 + 0.85 \ \mathrm{GeV}^{-2} \ (t-m_{K^{\ast}(892)}^2)\,.\label{eq: Kstartraj} 
\end{align}
In Refs.~\cite{reg_guidal, reg_guidal_2} it is argued that apart from the $K(494)$ and $K^{\ast}(892)$ trajectory exchanges, the Regge amplitude should also include the \emph{electric} contribution to the $s$-channel Born term: 
\begin{equation}
\begin{split}
\mathcal{M}_{Regge}\,(&\gamma\,p \rightarrow K^+\Sigma^0) =~ \mathcal{M}_{Regge}^K + \mathcal{M}_{Regge}^{K^{\ast}(892)} \quad \\ 
& + \mathcal{M}_{Feyn}^{p\ssst,\sst elec} \times \  \mathcal{P}_{Regge}^K \times \ (t-m_K^2)\,.
\label{eq: gauge_recipe}
\end{split}
\end{equation}
This is necessary because of the gauge-breaking nature of the $K^+$-exchange diagram. In a typical effective-Lagrangian framework, the Born terms $\mathcal{M}_{Feyn}^{p,K,\Sigma}$ in the $s$, $t$ and $u$ channels do not individually obey gauge invariance, whereas their sum does. Because the magnetic parts of the vertices ($\sim \sigma_{\mu\nu}q^{\nu}$) are gauge invariant by construction, only the electric parts ($\sim \gamma_{\mu}$) are of concern. It has been shown that the procedure of Eq.~(\ref{eq: gauge_recipe}) leads to a much-improved description of the high-energy $p(\gamma,K^+)\Lambda$ observables at low $\vert t \vert$~\cite{CorthalsL,Guidalthes, reg_guidal}.

When adopting interaction Lagrangians similar to those of Ref.~\cite{CorthalsL}, the high-energy amplitude for the $p(\gamma,K^+)\Sigma^0$ process contains only three free parameters:
\begin{equation}
g_{K^+\Sigma^0 p}\,, \quad G_{K^{\ast +}(892)}^{v,t} = \frac{e \, g_{{K^{\ast +}(892)}\, \Sigma^0 p}^{v,t}}{ 4 \pi} \ \kappa_{K^+{K^{\ast +}(892)}}\;,
\label{eq: bg_free_pars}
\end{equation}
with $g_{{K^{\ast +}(892)}\, \Sigma^0 p}^{v,t}$ 
the strong vector and tensor couplings to the $K^{\ast +}(892)$ vector meson trajectory. Assuming $SU(3)$-flavor symmetry, the strong $g_{K^+\Sigma^0 p}$ coupling constant can be related to the well-known $g_{\pi NN}$ coupling~\cite{deSwart,Mac87}. When allowing for a maximum deviation of 20\% from the exact $SU(3)$ value, the following range emerges:
\begin{equation}
0.9 ~ \le ~ \frac{g_{K^+\Sigma^0 p}}{\sqrt{4\pi}} ~ \le ~ 1.3\,.
\label{eq: su3}
\end{equation}
The $K^{\ast}(892)$ vector and tensor couplings are left entirely free. 

Apart from the three parameters of Eq.~(\ref{eq: bg_free_pars}), a choice between constant or rotating trajectory phases needs to be made. One may discriminate among the different alternatives by comparing the various model calculations with the results of high-energy measurements. Unfortunately, the published $p(\gamma,K^+)\Sigma^0$ data for $\omega_{lab} \gtrsim 4~\mathrm{GeV}$ are rather scarce. The relevant low-$\abs{t}$ data comprise 48 differential cross section points in total, at the selected energies $\omega_{lab}=5$, 8, 11 and $16~\mathrm{GeV}$~\cite{Boyarski69}, as well as 8 photon beam asymmetry points at $\omega_{lab} = 16$~GeV~\cite{Qui79}. No high-energy hyperon-polarization measurements have been performed for the $K\Sigma$ channels. 

In our previous work on $K\Lambda$ photoproduction~\cite{CorthalsL}, the recoil asymmetry $P$ was found to be particularly sensitive to the details of the Regge amplitude, much more so than the unpolarized cross section and photon beam asymmetry. The absence of recoil-polarization data for the $p(\gamma,K^+)\Sigma^0$ process constitutes a serious hindrance to constraining the various Regge-model parameters. Although a pure $t$-channel approach falls short of providing a complete quantitative description of the resonance-region data, the Regge model has been observed to reproduce all trends of the polarized and unpolarized $\gamma p \rightarrow K \Lambda/K^+\Sigma^0$ observables, including $P$~\cite{Guidalthes,reg_guidal,reg_guidal_2}. 
In view of these considerations, the procedure followed in this work amounts to discarding all Regge model variants which fail in reproducing the \emph{sign} of the recoil asymmetry in the resonance region.
Imposing this extra requirement reduces the number of possible model variants to four. They are classified in Table~\ref{tab: bg_specific} according to the sign of $G^{v,t}_{K^{\ast}(892)}$ and the phases of the $K(494)$ and $K^{\ast}(892)$ trajectories. The smaller values of $\chi^2$ as compared to what was found for the $K\Lambda$ channel~\cite{CorthalsL} can be attributed to the significantly larger error bars for the $K^+\Sigma^0$ high-energy cross sections. 
\begin{table}[t]
\caption{Fitted coupling constants for Regge model variants describing both the high-energy $p(\gamma,K^+)\Sigma^0$ data~\cite{Boyarski69,Qui79} and the sign of the recoil polarization in the resonance region~\cite{McNabb03}. The phase options for the $K(494)$ and $K^{\ast}(892)$ trajectories are listed in the second column. The last column mentions the attained $\chi^2$ value for the high-energy data. 
\label{tab: bg_specific}}
\begin{ruledtabular}
\begin{tabular}{l l  r r r r}
BG model & $K(494)$/$K^{\ast}(892)$ phase & $\frac{g_{K^+\Sigma^0 p}}{\sqrt{4 \pi}} $ & $G^v_{K^{\ast +}(892)}$ & $G^t_{K^{\ast +}(892)}$ & $\chi^2$ \hspace{-3.2pt} \rule[-10pt]{0pt}{21pt} \\
\hline 
\rule[-0pt]{0pt}{10pt}\hspace{-3.2pt}
1 & rot.~$K$, rot.~$K^{\ast}$ & 1.3  & 0.32 & 0.77 & 1.25\\
2 & rot.~$K$, rot.~$K^{\ast}$ & 1.3  & 0.33 & -0.86 & 1.28\\
3 & rot.~$K$, cst.~$K^{\ast}$ & 1.3 & -0.35 & 0.68 & 1.31\\
4 & rot.~$K$, cst.~$K^{\ast}$ & 1.3 & -0.32 & -0.87 & 1.27\rule[-3pt]{0pt}{2pt} \\
\end{tabular}
\end{ruledtabular}
\end{table}
\begin{figure}[b]
\begin{center}
\includegraphics[width=0.58\textwidth]{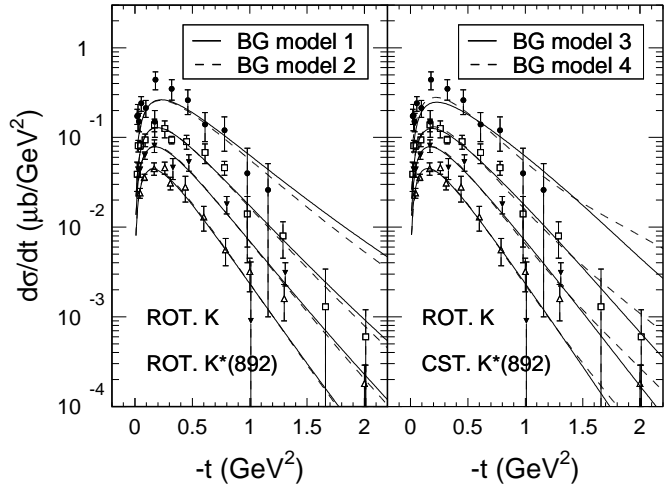}
\caption{Low-$t$ differential $p(\gamma,K^+)\Sigma^0$ cross sections at photon lab energies of 5~($\bullet$), 8~($\square$), 11~($\blacktriangledown$) and 16~($\bigtriangleup$) $\mathrm{GeV}$. The left panel corresponds to the Regge model variants with a rotating phase for the $K$ and $K^{\ast}$ trajectories. In the right panel, the model variants with a rotating $K$ and constant $K^{\ast}$ phase are shown.  The data are from Ref.~\cite{Boyarski69}.\label{fig: diffcs_highen}}
\end{center}
\end{figure}

\begin{figure}[t]
\begin{center}
\includegraphics[width=0.45\textwidth]{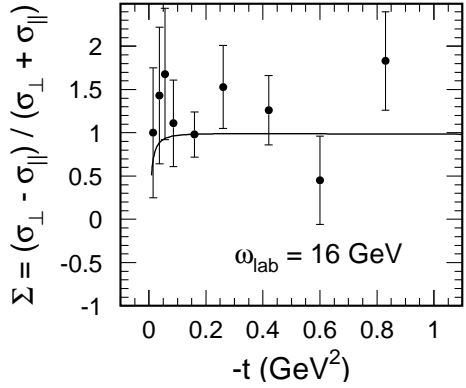}
\caption{Results for the forward-angle $p(\gamma,K^+)\Sigma^0$ photon beam asymmetry at $\omega_{lab}=16~\mathrm{GeV}$. The curves for the various background models are nearly indistinguishable. For the sake of clarity, only the result for model variant 1 is displayed. The data are from Ref.~\cite{Qui79}.\label{fig: phopol_highen}}
\end{center}
\end{figure}

\begin{figure}[t]
\begin{center}
\includegraphics[width=0.45\textwidth]{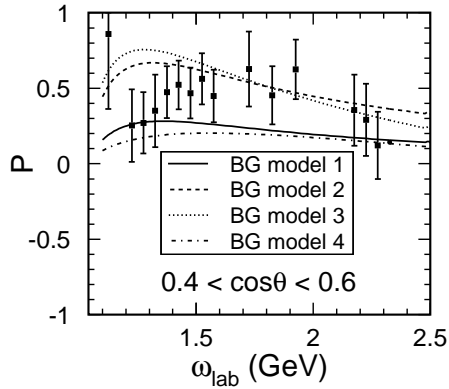}
\caption{Results for the $p(\gamma,K^+)\Sigma^0$ recoil asymmetry in the resonance region, for $0.4 < \cos\theta_{K}^{\ast} < 0.6$. The data are from Ref.~\cite{McNabb03}.\label{fig: recpol_resreg_bg}}
\end{center}
\end{figure}

A comparison between the calculated high-energy observables, resulting from the four Regge model variants of Table~\ref{tab: bg_specific}, and the data is shown in Figs.~\ref{fig: diffcs_highen}-\ref{fig: phopol_highen}. Figure~\ref{fig: recpol_resreg_bg} displays the recoil asymmetry in the resonance region for one representative $\cos\theta_K^{\ast}$ bin. As expected, the differential cross section (Fig.~\ref{fig: diffcs_highen}) and photon beam asymmetry (Fig.~\ref{fig: phopol_highen}) are rather insensitive to the choices made with respect to the trajectory phases and the signs of the coupling constants. On the other hand, the overall positive sign of the recoil asymmetry is only compatible with the four specific sign and phase combinations from Table~\ref{tab: bg_specific}. In particular, a strong correlation between the phase of the $K^{\ast}(892)$ trajectory and the sign of the corresponding vector coupling is observed. A rotating (constant) $K^{\ast}(892)$ phase requires a positive (negative) $G^v_{K^{\ast}}$ coupling.

It is clear from Table~\ref{tab: bg_specific} and Figs.~\ref{fig: diffcs_highen}-\ref{fig: phopol_highen} that the high-energy data do not allow to further discriminate between the retained Regge model variants, as all four provide a comparably good description. 

\subsection{Common analysis of $p(\gamma,K^+)\Sigma^0$ and $p(\gamma,K^0)\Sigma^+$}
\label{subsubsec:K0S+}

Due to the lack of $\gamma p \rightarrow K^0\Sigma^+$ data for $\omega_{lab} \gtrsim 4$ GeV, we have opted to constrain the Regge amplitude for this process against measurements performed at lower energies instead. This is deemed a feasible strategy, as the Regge model is known to provide very reasonable descriptions of the $K\Lambda$ and $K^+\Sigma^0$ photoproduction data in the resonance region~\cite{Guidalthes,reg_guidal,reg_guidal_2}. 

In principle, isospin arguments allow one to transform 
a reaction model for $\gamma p \rightarrow K^+ \Sigma^0$ into one for $\gamma p \rightarrow K^0 \Sigma^+$. By exploiting the fact that the $\Sigma^+$ and $\Sigma^0$ hyperons are members of an isotriplet, any coupling constant occurring in the $K^+\Sigma^0$ photoproduction amplitude can be converted into the corresponding $p(\gamma,K^0)\Sigma^+$ parameter. The strong coupling strengths are linked via $SU(2)$ Clebsch-Gordan coefficients, whereas for relating the electromagnetic couplings, experimental input in the form of $\Gamma_{K^{\ast}\rightarrow K\gamma}$ decay widths is required. The isospin relations used in this work can be found in the Appendix.

In practice, developing a common description for isospin-related channels is often less straightforward than one might infer from the preceding paragraph. Subtle interference effects might, for example, cause certain contributions to be masked in one channel, but strongly enhanced in the other. In fact, reconciling the $p(\gamma,K^+)\Sigma^0$ and $p(\gamma,K^0)\Sigma^+$ model predictions in the resonance region has proven challenging, as the measured $\Sigma^+$ cross-sections are considerably smaller than those for the $\Sigma^0$~\cite{Brad05,Glander04,Lawall05}. This observation is in apparent contradiction with the relation 
\begin{equation}
g_{K^0 \Sigma^+ p} = \sqrt{2} \; g_{K^+ \Sigma^0 p}\,
\label{eq: strongccrel_K}
\end{equation}
(see Appendix), with similar expressions holding when a $N^{\ast}$, $K^{\ast}$ or $Y^{\ast}$ resonance is involved at the vertex. 

In isobar models, this difficulty is often circumvented by strongly reducing the $g_{K\Sigma p}$ coupling in both channels (thus disregarding the $SU(3)$ constraints of Eq.~(\ref{eq: su3})), and/or by carefully counterbalancing the superfluous strength in the $K^0\Sigma^+$ channel through destructive interferences induced by other contributions~\cite{Stijn_sigma,MaBe95}. It shall be demonstrated that, in the context of the RPR approach, this issue can be elegantly resolved at the level of the \emph{background} terms. 
 
It will become clear that the $p(\gamma,K^+)\Sigma^0$ Regge model variants proposed in Sec.~\ref{subsubsec:K+S0} cannot be readily extended to the $K^0\Sigma^+$ channel. Since the $\gamma K K$ vertex is proportional to the kaon charge, the $K$-trajectory exchange diagram, as well as the accompanying gauge-restoring $s$-channel electric Born term, do not contribute to the $K^0\Sigma^+$ amplitude. 
Therefore, the equivalent of Eq.~(\ref{eq: gauge_recipe}) in this channel simply reads:
\begin{equation}
\mathcal{M}_{Regge}\,(\gamma\,p \rightarrow K^0\Sigma^+) =~ \mathcal{M}_{Regge}^{K^{\ast}(892)} \,.
\label{eq: ampli_kps0}
\end{equation}

\begin{figure}[ht]
\begin{center}
\includegraphics[width=0.45\textwidth]{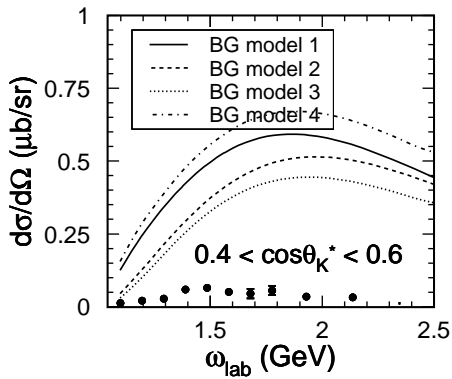}
\caption{Results for the $p(\gamma,K^0)\Sigma^+$ differential cross section in the resonance region for $0.4 < \cos\theta_{K}^{\ast} < 0.6$, obtained by converting the model parameters from Table~\ref{tab: bg_specific} to the $K^0\Sigma^+$ channel. The data are from Ref.~\cite{Lawall05}.\label{fig: diff_iso3_iso2pred}}
\end{center}
\end{figure}

Figure~\ref{fig: diff_iso3_iso2pred} displays the predictions for the $p(\gamma,K^0)\Sigma^+$ differential cross section for one particular $\cos\theta_{K}^{\ast}$ bin in the resonance region, using the above-mentioned form for the amplitude. The $G^{v,t}_{K^{\ast 0}(892)}$ couplings have been determined through the isospin relations from the Appendix,  starting from the fitted values listed in Table~\ref{tab: bg_specific}. It is instantly clear that the model parameters determined from the high-energy $\gamma p \rightarrow K^+ \Sigma^0$ data, when converted to the $K^0\Sigma^+$ channel, result in cross sections that overshoot experimental data by a factor of 10.
Thus, an amplitude of the type of Eq.~(\ref{eq: ampli_kps0}) apparently does not suffice to provide a reasonable description of the $\gamma p \rightarrow K^0 \Sigma^+$ data. 
In this repect, we deem it relevant to mention the $K^0\Sigma^+$ total cross-section result obtained by Guidal and Vanderhaeghen~\cite{Guidal_elec} by means of Eq.~(\ref{eq: ampli_kps0}). As confirmed to us by the authors~\cite{Guidal_pc}, the curves shown in Fig. 3 of their article do not take into account the isospin factor of $\sqrt{2}$, relating the strong $g_{K^+\Sigma^0 p}$ and $g_{K^0\Sigma^+ p}$ couplings [Eq.~(\ref{eq: strongccrel_K})]. Inclusion of this factor would increase the quoted cross section by a factor of two, considerably worsening the quality of agreement with the data.

A parallel can be drawn between the Regge descriptions of photoinduced kaon and pion production. Indeed, in Refs.~\cite{Guidalthes,reg_guidal,reg_guidal_2}, Guidal and Vanderhaeghen modelled the charged-$\pi$ photoproduction channels through $\pi$ and $\rho$ trajectory exchanges. In $\pi^0$ production, on the other hand, an $\omega$ trajectory was introduced to compensate for the vanishing $\pi$-exchange diagram. Similarly, in the absence of a $K(494)$ contribution to the $K^0\Sigma^+$ amplitude, a higher-mass trajectory may become important in this channel, serving to counterbalance the $K^{\ast}(892)$ strength.

It can be intuitively understood that the strong destructive interference needed to reduce the predicted cross sections to the level of the data (see Fig.~\ref{fig: diff_iso3_iso2pred}) can be efficiently realized when the added contribution exhibits an angular distribution comparable to that of the $K^{\ast}(892)$-exchange diagram. This implies that a natural-parity particle should be involved. A second $K^{\ast}$ trajectory is likely to realize the required effect. As it turns out, the PDG tables hint at the presence of such a trajectory, with the $K^{\ast}(1410)$ vector particle as first materialization and the $K^{\ast}_2(1980)$ as a probable second member. However, whereas the meson trajectories tend to possess a more or less universal slope, the slope of this experimental $K^{\ast}(1410)$ trajectory is significantly smaller than those of the well-known $K(494)$ and $K^{\ast}(892)$ trajectories, i.e.~$0.53$ GeV$^{-2}$ as compared to 0.7 and 0.85 GeV$^{-2}$.

\begin{figure}[t]
\begin{center}
\includegraphics[width=0.45\textwidth]{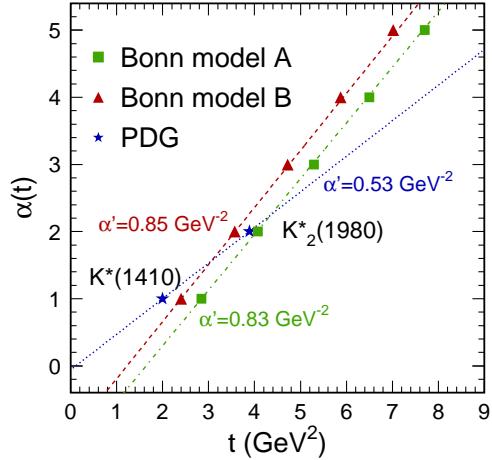}
\caption{(Color online) Comparison between the experimental $K^{\ast}(1410)$ trajectory and the Bonn-model predictions~\cite{bonn}. The experimental meson masses are from the Particle Data Group~\cite{PDG04}.}
\label{fig: newkst-traj_bonn}
\end{center}
\end{figure}

As the properties of the $K^{\ast}(1410)$ trajectory cannot be put on solid grounds with the available experimental information, we also turned our attention to the predictions of a constituent-quark model (CQM) calculation of the kaon spectrum. The Lorentz covariant quark model developed by the Bonn group~\cite{bonn} provides a satisfactory description of the light meson masses and decay properties. Figure~\ref{fig: newkst-traj_bonn} displays the results of the calculations using two different options ($A$ and $B$) for the Dirac structure of the confinement potential. 
After selecting from the predicted spectra the states most likely to correspond to the $K^{\ast}(1410)$ and $K^{\ast}_2(1980)$ resonances, and supplementing these with a set of suitable higher-spin states, a linear relation presents itself. 

The slopes of the theoretical and experimental trajectories clearly differ. Strikingly, however, the two calculated curves have practically identical slopes, which are also perfectly compatible with those of the $K(494)$ and $K^{\ast}(892)$ trajectories [Eqs.~(\ref{eq: Ktraj})-(\ref{eq: Kstartraj})]. The calculated masses and spins of the members of the $K^{\ast}(1410)$ trajectory are nearly perfectly fitted by a linear curve. 

We have opted to use the calculated value of 0.83 GeV$^{-2}$, corresponding to the Bonn-model variant~$A$, leading to a trajectory of the form
\begin{equation}
\alpha_{K^{\ast}(1410)}(t) = 1 + 0.83 \ \mathrm{GeV}^{-2} \ (t-m_{K^{\ast}(1410),\mathrm{PDG}}^2)\,,\label{eq: 2ndKstartraj} 
\end{equation}
with $m_{K^{\ast}(1410),\mathrm{PDG}} =  1414$ MeV. The corresponding Regge propagator takes on the form of Eq.~(\ref{eq: reggeprop_Kstar}). Again, the trajectory phase may either be constant or rotating. 

After adding the $K^{\ast}(1410)$-trajectory exchange diagram, the $\gamma p \rightarrow K^0 \Sigma^+$ Regge amplitude is given by
\begin{equation}
\mathcal{M}_{Regge}\,(\gamma\,p \rightarrow K^0\Sigma^+) =~ \mathcal{M}_{Regge}^{K^{\ast}(892)} + \mathcal{M}_{Regge}^{K^{\ast}(1410)}
\label{eq: ampli_kps0_right}
\end{equation}
and the number of model parameters is increased by two. Contrary to the $K^{\ast}(892)$ parameters, which are constrained by the high-energy $p(\gamma,K^+)\Sigma^0$ data, the $K^{\ast}(1410)$ vector and tensor couplings remain as yet unknown, as does the matching trajectory phase. In the absence of high-energy data for the $K^0\Sigma^+$ channel, we fix the $K^{\ast}(1410)$ parameters through a fit to the forward-angle ($\cos\theta > 0$) part of the $p(\gamma,K^0)\Sigma^+$ differential cross-section data for the resonance region. 

\begin{table}[ht]
\caption{Extracted $K^{\ast 0}(1410)$ parameters for each of the model variants from Table~\ref{tab: bg_specific}. The trajectory phase options are given in the second column, while the last column shows the attained $\chi^2$ value when comparing to the resonance-region $p(\gamma,K^0)\Sigma^+$ cross-section data. 
\label{tab: bg_specific_k0sp}}
\begin{ruledtabular}
\begin{tabular}{l l  r r r}
BG mod. & $K^{\ast}(892)$/$K^{\ast}(1410)$ phase & $G^v_{K^{\ast 0}(1410)}$ & $G^t_{K^{\ast 0}(1410)}$ & $\chi^2$ \hspace{-3.2pt} \rule[-10pt]{0pt}{21pt} \\
\hline 
\rule[-0pt]{0pt}{10pt}\hspace{-3.2pt}
1 & rot.~$K^{\ast}(892)$, rot.~$K^{\ast}(1410)$ &  -3.0 & -5.0 & 11.8 \\
2 & rot.~$K^{\ast}(892)$, rot.~$K^{\ast}(1410)$ & -3.4 & 4.5 & 8.3 \\
3 & cst.~$K^{\ast}(892)$, cst.~$K^{\ast}(1410)$ & -3.1 & 6.1 & 10.5 \\
4 & cst.~$K^{\ast}(892)$, cst.~$K^{\ast}(1410)$ & -2.9 & -6.3 & \rule[-3pt]{0pt}{2pt} 10.2 \\
\end{tabular}
\end{ruledtabular}
\end{table}
\begin{figure}[b]
\begin{center}
\includegraphics[width=0.4\textwidth]{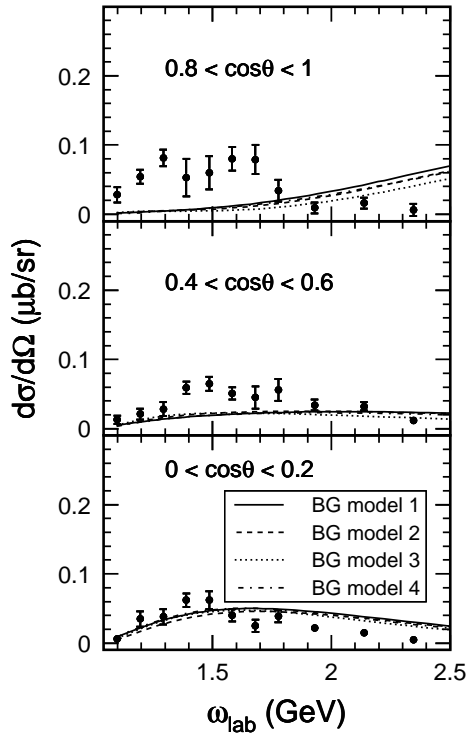}
\caption{Results for the $p(\gamma,K^0)\Sigma^+$ differential cross section in the resonance region. The data are from Ref.~\cite{Lawall05}.\label{fig: diff_iso3_bgonly}} 
\end{center}
\end{figure}
Table~\ref{tab: bg_specific_k0sp} displays the extracted $K^{\ast}(1410)$ parameters for each of the background models proposed in Sec.~\ref{subsubsec:K+S0}. The values of the $G^{v,t}_{K^{\ast 0}(892)}$ couplings can be found from the $G^{v,t}_{K^{\ast +}(892)}$ values (Table~\ref{tab: bg_specific}) by applying the relations from the Appendix. It is clear from Table~\ref{tab: bg_specific_k0sp} that the $K^{\ast}(892)$ and $K^{\ast}(1410)$ trajectory phases are strongly coupled. It can be intuitively understood that destructive interference is strongest when the same phase choice is adopted for both trajectories. Because of its larger mass, the Regge propagator for the $K^{\ast}(1410)$ is smaller than the $K^{\ast}(892)$ one, hence a larger coupling constant is needed to produce contributions of similar magnitude.

The results for the $p(\gamma,K^0)\Sigma^+$ differential cross section in the resonance region are shown in Fig.~\ref{fig: diff_iso3_bgonly} for three bins in the forward hemisphere of $\cos\theta_K^{\ast}$. Apart from the slight rise with energy at $\cos\theta_K^{\ast} \approx 0.9$, the order of magnitude of the experimental curves is now reasonably well-matched by the calculations. While this channel appears to be background-dominated, some resonance dynamics are clearly missing in the $\omega_{lab} \lesssim 1.7$ GeV region, especially at the more forward angles. This will be remedied in the following section.

\begin{figure}[t]
\begin{center}
\includegraphics[width=0.37\textwidth]{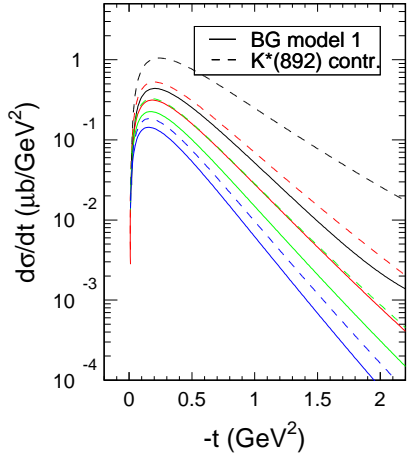}
\caption{(Color online) Predictions for the low-$t$ differential $p(\gamma,K^0)\Sigma^+$ cross sections at photon lab energies of 5, 8, 11 and 16 $\mathrm{GeV}$ (the highest energy corresponding to the smallest cross-section) using the background model variant 1. The full curves represent the total amplitude, whereas the dashed curves show the contribution of the $K^{\ast}(892)$ trajectory.\label{fig: diff_highen_iso3_pred}}
\end{center}
\end{figure}

The inclusion of the $K^{\ast}(1410)$ trajectory in the $K^0\Sigma^+$ channel also affects the high-energy observables. Figure~\ref{fig: diff_highen_iso3_pred} displays a prediction for the $p(\gamma,K^0)\Sigma^+$ differential cross section, using the Regge model variant~1,  at photon lab energies of 5, 8, 11 and 16 GeV. The other three model variants result in a comparable behavior for the cross section. When comparing Fig.~\ref{fig: diff_highen_iso3_pred} with Fig.~\ref{fig: diffcs_highen}, it is clear that the Regge amplitude of Eq.~(\ref{eq: ampli_kps0_right}), incorporating both $K^{\ast}$ trajectories (full lines), produces cross sections of the same order of magnitude as those for the $\gamma p \rightarrow K^+\Sigma^0$ process. On the other hand, use of Eq.~(\ref{eq: ampli_kps0}), accounting for the $K^{\ast}(892)$ trajectory exchange only (dashed lines), leads to cross sections that are higher by a factor of 2 up to 10, depending on the energy. It can be seen that the relative importance of the $K^{\ast}(1410)$ contribution diminishes with increasing photon energy. We wish to stress that Fig.~\ref{fig: diff_highen_iso3_pred} shows a prediction for the $p(\gamma,K^0)\Sigma^+$ cross section at high energies, obtained with background parameters constrained by the \emph{resonance-region} data. A high-energy measurement performed for this reaction channel would prove extremely useful in putting these predictions to a stringent test. 

\section{$\bm{K\Sigma}$ photoproduction in the resonance region}
\label{subsec: results res}

With a view to weighing the importance of the various $N^{\ast}$ and $\Delta^{\ast}$ contributions, we perform not merely one, but a series of fits to the resonance-region data. The resonances are added one at a time, allowing one to check the impact of each candidate on the attained value of $\chi^2$. 

We analyze the high-precision $p(\gamma,K^+)\Sigma^0$ data from CLAS, comprising an extensive set of unpolarized cross sections and hyperon polarizations~\cite{McNabb03,Brad05}. Photon-beam asymmetry data for the second and third resonance regions, taken specifically at forward kaon angles, have been provided by the LEPS collaboration~\cite{Zegers03}. In addition, the GRAAL collaboration has been involved in beam- and recoil- polarization measurements in the first resonance region over an extensive angular range~\cite{Lleres06}. For $p(\gamma,K^0)\Sigma^+$, the differential cross-section and recoil-asymmetry data provided by the SAPHIR collaboration are employed~\cite{Lawall05}. 

As this work hinges on $t$-channel reggeization, only the forward-angle portion of the various datasets is considered for fitting purposes. For the copious CLAS data the restriction $\cos\theta_K^{\ast}>0.35$ is imposed, whereas for the LEPS, GRAAL and SAPHIR data the $\cos\theta_K^{\ast}>0.0$ part is taken into account. We opted for this particular selection of data points in order to be consistent with our previous RPR study of the $K^+\Lambda$ channel~\cite{CorthalsL}. This leaves in total 618 data points with which to adjust the model parameters. The quoted number includes 435 differential cross sections, 53 recoil asymmetries (49 from CLAS and 4 from GRAAL) and 66 photon beam asymmetries (45 from LEPS and 21 from GRAAL) for the $K^+\Sigma^0$ channel, and for the $K^0\Sigma^+$ channel 60 differential cross sections plus 4 recoil asymmetry points. It is worth stressing that in our RPR approach, the only parameters that remain to be fitted to the resonance-region data are the resonance couplings and the cutoff $\Lambda_{res}$ for the strong resonance form factors. Moreover, for the masses and widths of the known resonances we assume the PDG values~\cite{PDG04} instead of treating them as free parameters as is often done. 

The complex issue of minimizing $\chi^2$ is tackled using a combination of a simulated annealing algorithm (SAA)~\cite{Simann,Stijnthes} and the CERN MINUIT~\cite{Minuit} package. Starting points for minimization are obtained from the SAA, which was designed to produce parameters near the global minimum of the $\chi^2$ surface. The parameters provided by the SAA are then fed into MINUIT in order to pinpoint the location of each minimum more precisely, and obtain an error matrix for the fitted parameters. In previous studies~\cite{GApaper03,GApaper04} we have used a genetic algorithm and many calculations for each model variant to explore parameter space more fully. In the present case we have a large number of model variants, making a more exhaustive investigation too cumbersome.

In the literature, there is some diversity of opinion about the
resonant contributions to the $p(\gamma,K)\Sigma$ channels. Most of the published models are based
on the SAPHIR data released in the late nineties, whereas the few
analyses that employ the most recent datasets appear to lead to different conclusions. Since the $K^+\Sigma^0$ photoproduction data, unlike the $K^+\Lambda$ ones, do not exhibit an explicit resonant structure, it was long deemed unnecessary to introduce any ``missing'' states in the $\Sigma$ channels~\cite{Stijn_sigma,Mosel02_pho,US05}. The new data, however, are characterized by significantly reduced error bars, so that a detailed analysis may reveal effects previously clouded by experimental uncertainty. It has been shown that the $K^0\Sigma^+$ observables in particular may point to a second $S_{11}$ resonance~\cite{Lawall05}, indications of which have also been reported for the $K\Lambda$ channel~\cite{Diaz05}. On the other hand, the recent analysis of the $K^+\Sigma^0$ and $K^0\Sigma^+$ photoproduction channels by Sarantsev et al.~\cite{Sara05} calls for the inclusion of missing
$D_{13}(1870)$, $D_{13}(2170)$ and $P_{11}(1840)$ states.

As each resonant contribution implies the introduction of at least one parameter, we aim at keeping them at a strict minimum. We therefore consider only resonances with spin $J \leq$ 3/2 and a mass below 2 GeV. We further limit ourselves to the established PDG resonances with a star classification of two or higher. No ``missing'' states are included at this point.

\begin{table}[b]
\caption{Combinations of resonances used in the calculations. The
  second column mentions the number of free parameters (NFP) for each
  model variant, not including the background couplings. 
The last four columns list the values of $\chi^2$ attained by combining each resonance set (RS) with background options 1 through 4 from Table~\ref{tab: bg_specific}, and adjusting the resonance parameters to the resonance-region data. \label{tab: allRPR}}
\begin{ruledtabular}
\begin{tabular}{c c c c c c c c c c c c}
RS & NFP & $N^{\ast}$ & $P_{13}(1900)$ & $S_{31}(1900)$ & $P_{31}(1910)$ & $D_{33}(1700)$ & $P_{33}(1920)$ & \multicolumn{4}{c}{$\chi^2$ for BG mod.}\\ 
& & core  & & & & & & \ 1 \ & \ 2 \ & \ 3 \ & \ 4 \ \\
\hline
A & 10 & $\bigstar$ & -- & $\bigstar$ & $\bigstar$ & -- & -- & 10.4 & 6.0 & 9.0 & 3.8 \\
B & 15 & $\bigstar$ & $\bigstar$ & $\bigstar$ & $\bigstar$ & -- & -- & 4.4 & 4.0 & 4.4 & 3.4 \\ 
C & 20 & $\bigstar$ & $\bigstar$ & $\bigstar$ & $\bigstar$ & $\bigstar$ & -- & 3.5 & 3.6 & 2.2 & 2.1 \\ 
D & 19 & $\bigstar$ & $\bigstar$ & -- & $\bigstar$ & $\bigstar$ & -- & 3.5 & 3.7 & 2.4 & 2.3 \\ 
E & 19 & $\bigstar$ & $\bigstar$ & $\bigstar$ & -- & $\bigstar$ & -- & 3.7 & 3.8 & 3.1 & 2.4 \\ 
F & 20 & $\bigstar$ & $\bigstar$ & $\bigstar$ & $\bigstar$ & -- & $\bigstar$ & 3.5 & 3.7 & 3.2 & 2.6 \\ 
G & 19 & $\bigstar$ & $\bigstar$ & -- & $\bigstar$ & -- & $\bigstar$ & 3.6 & 3.7 & 3.6 & 2.9 \\ 
H & 19 & $\bigstar$ & $\bigstar$ & $\bigstar$ & -- & -- & $\bigstar$ & 3.7 & 4.1 & 3.3 & 2.7 \\ 
I & 25 & $\bigstar$ & $\bigstar$ & $\bigstar$ & $\bigstar$ & $\bigstar$ & $\bigstar$ & 3.5 & 2.8 & 2.0 & 2.0 \\ 
\end{tabular}
\end{ruledtabular}
\end{table}

Table~\ref{tab: allRPR} gathers the combinations of nucleon and
$\Delta$ resonances assumed in the various calculations. All resonance
sets (RS) A through I are combined with each of the four background
options constructed in Sec.~\ref{subsec: results bg}. This amounts to
thirty-six RPR model variants. The listed values of $\chi^2$ stem from a comparison of the computed
$p(\gamma,K^+)\Sigma^0$ and $p(\gamma,K^0)\Sigma^+$ observables with the resonance-region data of Refs.~\cite{McNabb03, Brad05, Zegers03, Lleres06,Lawall05}. The simplest resonance set, RS A, corresponds to the standard combination of states assumed in most of the early isobar calculations~\cite{Kaon-maid,Stijn_sigma}. As in the $K\Lambda$ channel, the ``core'' $N^{\ast}$ set consists of the $S_{11}(1650)$, $P_{11}(1710)$ and $P_{13}(1720)$ resonances. Two $\Delta^{\ast}$ states, $S_{31}(1900)$ and $P_{31}(1910)$, each having spin 1/2 and thus involving one extra free parameter, are also included in RS A. We further consider three additional spin-3/2 resonances: $P_{13}(1900)$, $D_{33}(1700)$ and $P_{33}(1920)$.

From Table~\ref{tab: allRPR}, a number of trends may readily be spotted. Only background model (BGM) 1 fails to produce a $\chi^2$ smaller than 3.5 in combination with any of the resonance sets. BGM 2 leads to somewhat better results, although 25 parameters are required to reduce $\chi^2$ below 3.6. Both of the above-mentioned models assume rotating phases for all Regge trajectories, so one may conclude that this choice - though adequate for the high-energy description - is less suitable for the resonance region. The models assuming constant $K^{\ast}(892)$ and $K^{\ast}(1410)$ phases along with a rotating $K(494)$ phase, i.e.~BGMs 3 and 4, perform considerably better. Although the minimal value of $\chi^2 =2.0$ is identical for both model variants, BGM 4 exhibits significantly less need for the inclusion of additional resonances than BGM 3. This is evident when comparing the $\chi^2$ values found for the resonance sets with the smallest (RS A) and largest (RS I) number of free parameters. Note also that all BGM-4 variants with more than 15 free parameters have a $\chi^2$ below 3.0, contrary to BGM 3, for which only three of the resonance sets (C, D and I) perform this well. It can be seen that BGM 4 consistently provides the best fit to the data, irrespective of which resonance contributions are added. This leads to the conclusion that the resonance-region data prefer a negative sign for the $K^{\ast}(892)$ tensor coupling, as in BGM 4, rather than a positive one as is the case for BGM 3 (see Table~\ref{tab: bg_specific}).

Table~\ref{tab: allRPR} also prompts a number of conclusions regarding the resonant structure of the amplitudes. Comparing RS A and B shows that adding the $P_{13}(1900)$ state significantly improves $\chi^2$ for all four Regge model variants. On the other hand, the conclusions with regard to the $\Delta$ resonances clearly depend on the background choice. The general trends for the ``preferred'' background models 3 and 4 are largely comparable, however. Including the $D_{33}(1700)$ state considerably reduces $\chi^2$, in contrast to $P_{33}(1920)$, which has a fairly limited impact on the quality of the fit (compare RS C to F, D to G, and E to H). Removing either $S_{31}(1900)$ or $P_{31}(1910)$ does not spoil the agreement with the data, indicating that only a single spin-1/2 $\Delta$ resonance is required, the parity of which remains unclear.

\begin{table}[t]
\caption{RPR model variants providing the best common description of the $p(\gamma,K^+)\Sigma^0$ and $p(\gamma,K^0)\Sigma^+$ data from threshold up to 16 GeV in photon energy. The Regge background model (BM) and resonance set (RS) are given, using the numbering from Tables~\ref{tab: bg_specific} and \ref{tab: allRPR}. 
Also listed are the partial $\chi^2$ values for the $K^+\Sigma^0$ and $K^0\Sigma^+$ channels, as well as the total $\chi^2$. \label{tab: res_specific}}
\begin{ruledtabular}
\begin{tabular}{l c c c c r}
RPR & BG mod. & RS & $\chi^2_{K^+\Sigma^0}$ & $\chi^2_{K^0\Sigma^+}$ & \ $\chi^2$\ \\ 
\hline
RPR-3 & 3 & I & 1.8 & 3.8 & 2.0 \\
RPR-4 & 4 & I & 1.9 & 2.6 & 2.0 \\
\end{tabular}
\end{ruledtabular}
\end{table}

\begin{figure*}[b]
\begin{center}
\includegraphics[width=0.52\textwidth]{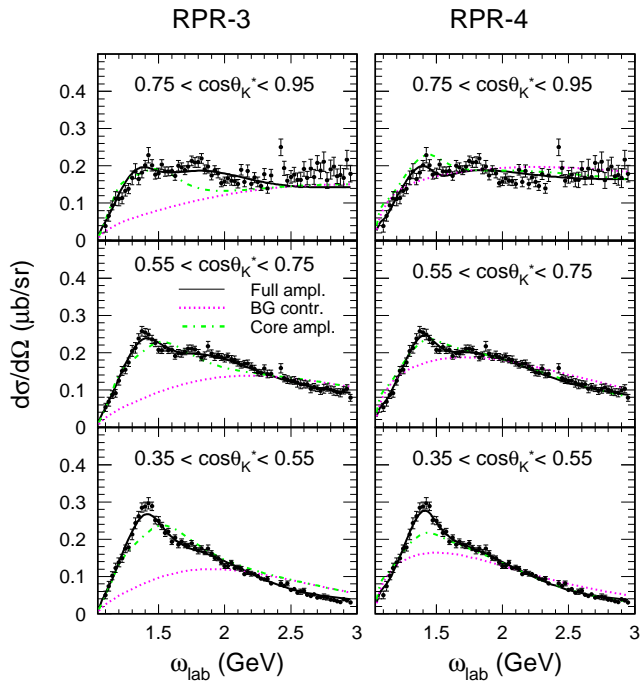}
\caption{(Color online) Energy dependence of the differential $p(\gamma,K^+)\Sigma^0$ cross sections in the resonance region, for a number of representative bins in $\cos\theta_{K}^{\ast}$. The full curves represent the complete result, the dotted curves show the contribution of the Reggeized background (BG) amplitude, whereas the dot-dashed curves correspond to RPR model variants containing only the ``core'' resonance set A from Table~\ref{tab: allRPR} (see text). The data are from CLAS~\cite{Brad05}.
\label{fig: diffcs_2_lowen} }
\end{center}
\end{figure*}

Judging by the $\chi^2$ values from Table~\ref{tab: allRPR}, the two RPR model variants providing the best common description of the high- and low-energy $p(\gamma,K)\Sigma$ observables are those assuming background options 3 and 4, combined with the most complete resonance set, RS I. These models will be referred to as RPR-3 and RPR-4, respectively. The specifications for both are summarized in Table~\ref{tab: res_specific}. 

\begin{figure*}[b]
\begin{center}
\includegraphics[width=0.52\textwidth, clip]{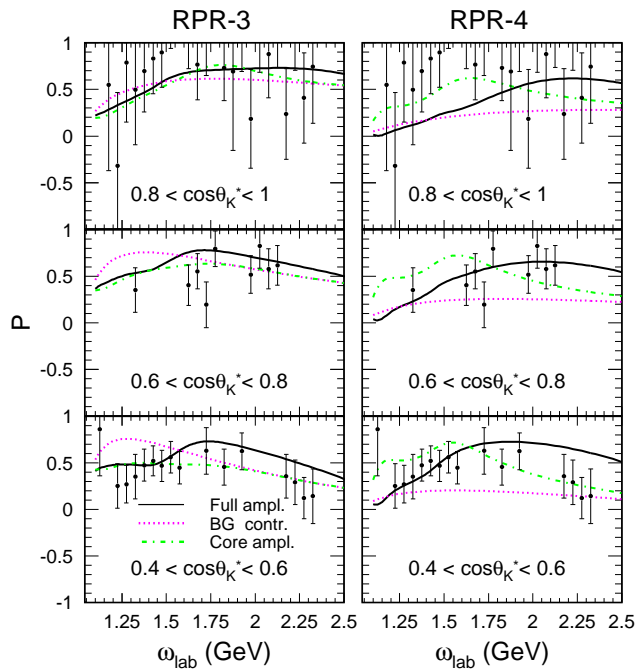}
\caption{(Color online) Energy dependence of the $p(\gamma,K^+)\Sigma^0$ recoil polarization for those bins of $\cos\theta_{K}^{\ast}$ considered in the fitting procedure. Line conventions are as in Fig.~\ref{fig: diffcs_2_lowen}. The data are from CLAS~\cite{McNabb03}. \label{fig: recpol_2_lowen}}
\end{center}
\end{figure*}

\begin{figure*}[t]
\begin{center}
\includegraphics[width=0.52\textwidth, clip]{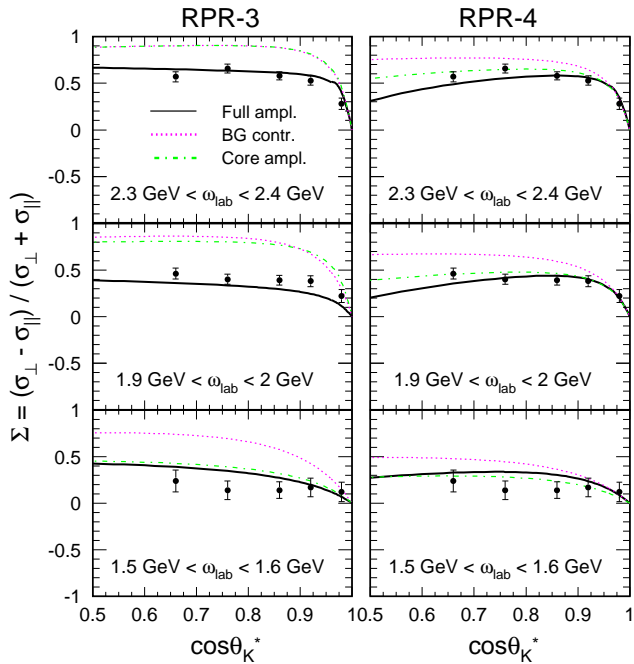}
\caption{(Color online) Results for the forward-angle $p(\gamma,K^+)\Sigma^0$ photon beam asymmetry, for three representative bins of $\omega_{lab}$, corresponding to center-of-mass energy bins $ 2.28~\mathrm{GeV}<W<2.32~\mathrm{GeV}$, $2.11~\mathrm{GeV}<W<2.15~\mathrm{GeV}$, and $1.92~\mathrm{GeV}<W<1.97~\mathrm{GeV}$. Line conventions are as in Fig.~\ref{fig: diffcs_2_lowen}. The data are from LEPS~\cite{Zegers03}. \label{fig: phopol_lowen}}
\end{center}
\end{figure*}

\begin{figure*}[t]
\begin{center}
\includegraphics[width=0.52\textwidth]{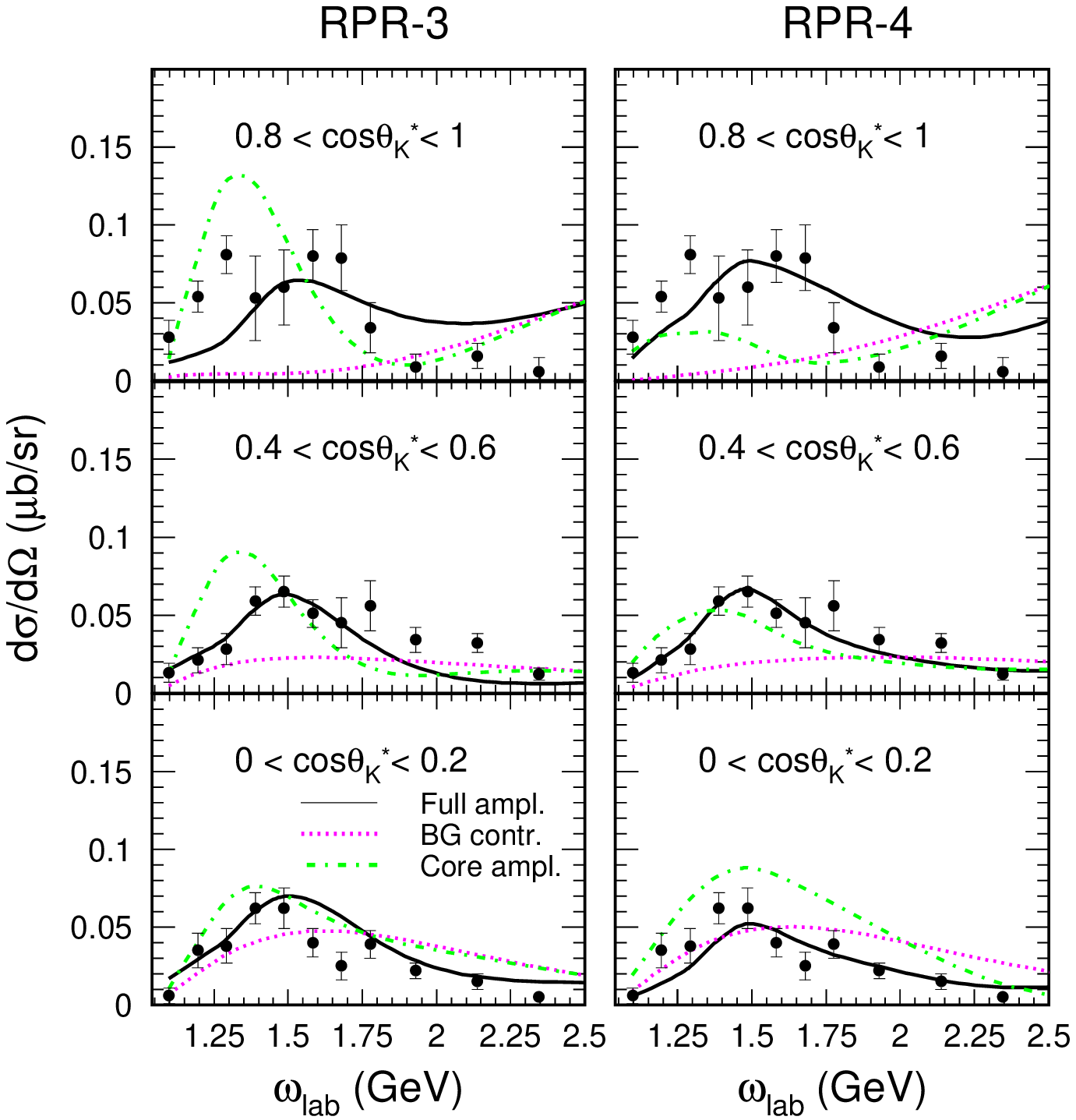}
\caption{(Color online) Energy dependence of the differential $p(\gamma,K^0)\Sigma^+$ cross sections in the resonance region, for a number of representative bins in $\cos\theta_{K}^{\ast}$. Line conventions are as in Fig.~\ref{fig: diffcs_2_lowen}. The data are from SAPHIR~\cite{Lawall05}.
\label{fig: diffcs_3_lowen} }
\end{center}
\end{figure*}

\begin{figure*}[t]
\begin{center}
\includegraphics[width=0.52\textwidth]{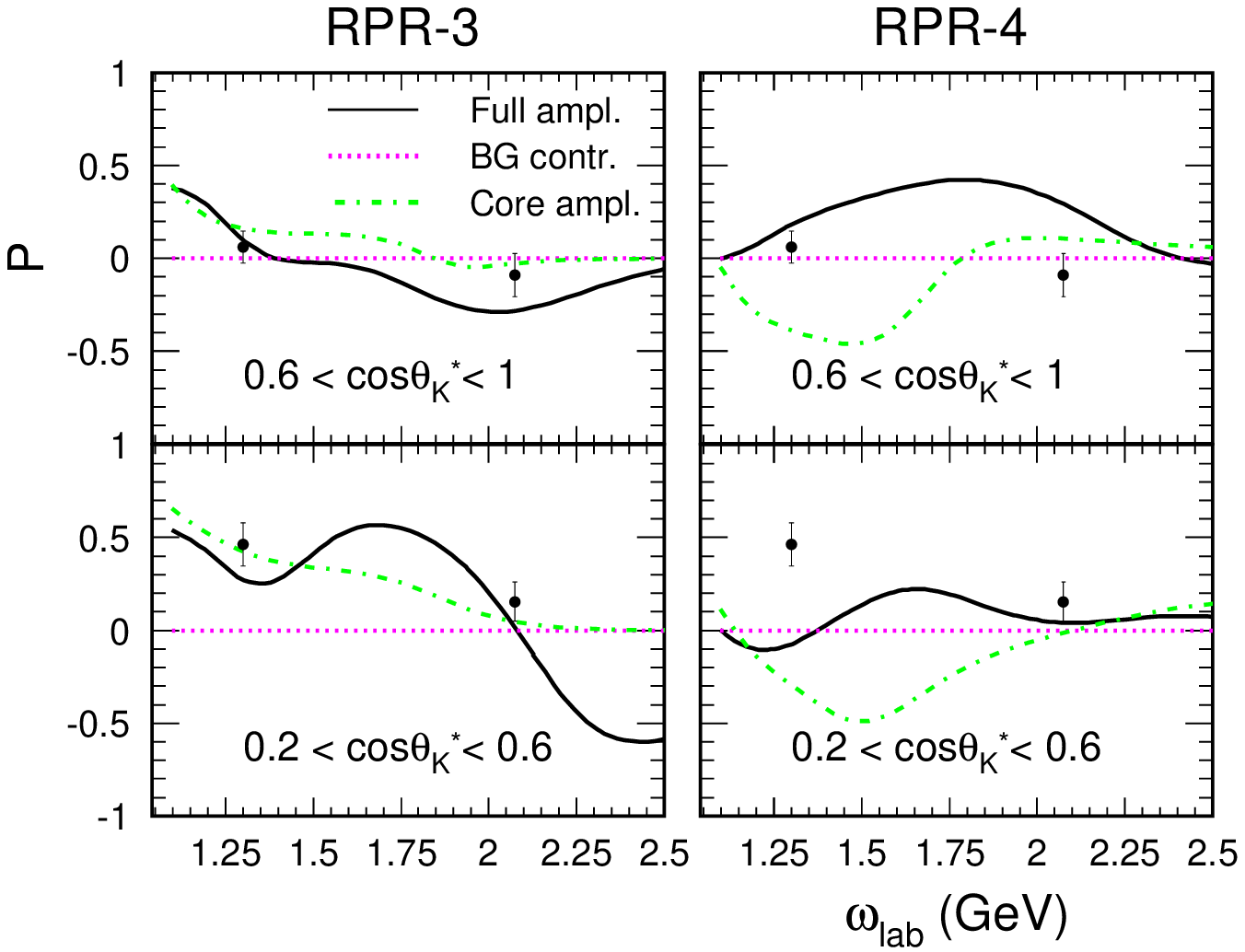}
\caption{(Color online) Energy dependence of the $p(\gamma,K^0)\Sigma^+$ recoil polarization for those bins of $\cos\theta_{K}^{\ast}$ considered in the fitting procedure. Line conventions are as in Fig.~\ref{fig: diffcs_2_lowen}. The data are from SAPHIR~\cite{Lawall05}.
\label{fig: recpol_3_lowen} }
\end{center}
\end{figure*}

The results of the RPR-3 and RPR-4 calculations for the various $p(\gamma,K)\Sigma$ observables are compared to the world data in Figs.~\ref{fig: diffcs_2_lowen}-\ref{fig: recpol_3_lowen}. The curves indicated as ``BG'' correspond to the background contributions to the full RPR amplitudes. Also displayed are the results for two alternative RPR model variants, consisting of the ``core'' resonance set A from Table~\ref{tab: allRPR} in combination with background model variants 3 and 4, respectively. 

Figure~\ref{fig: diffcs_2_lowen} shows the $p(\gamma,K^+)\Sigma^0$
differential cross section as a function of $\omega_{lab}$. Both RPR-3 and RPR-4 succeed remarkably well in reproducing this observable, including the subtle ``shoulder'' in the energy dependence
at $\omega_{lab} \approx 1.75$ GeV ($W \approx 2.05$ GeV), which is
likely to arise from destructive interference of the background with
resonances in the 1900-GeV mass range. The ``core'' models, containing only lower-mass resonances, clearly
fall short on this account. In addition, they seriously underestimate
the value of the cross-section maximum at the more backward kaon
angles. Similar to the $K\Lambda$ case, the Regge model
produces smooth curves. Towards the highest $\omega_{lab}$ measured by
CLAS, it describes the unpolarized data without the inclusion of any resonant
diagrams. For $\omega_{lab} \lesssim$ 2 GeV ($W \lesssim$ 2.15
GeV), $s$-channel contributions are obviously required. 

The computed recoil polarization $P$ and photon beam asymmetry
$\Sigma$ are shown in Figs.~\ref{fig: recpol_2_lowen} and \ref{fig: phopol_lowen}. Both observables are well reproduced by RPR-3 and RPR-4. Again, the Regge contribution in itself provides a good approximation of the experimental hyperon polarization. This justifies the choice to constrain the Regge model variants by requiring them to predict the correct sign for $P$ in the resonance region. In their description of the recoil asymmetry, the ``core'' models perform comparably to RPR-3 and RPR-4, indicating that the size of the error bars for this observable hampers the extraction of information on the underlying resonance structure. While the Regge and core amplitudes reproduce the sign of 
the photon beam asymmetry, its magnitude and energy dependence can
only be explained by a reaction model containing a sufficiently large number of
resonances. The impact of the resonant part of the amplitude on $P$
and $\Sigma$ persists up to significantly higher energies than was the
case for the unpolarized cross section. 

The results for the $p(\gamma,K^0)\Sigma^+$ channel are shown in Figs.~\ref{fig: diffcs_3_lowen} and \ref{fig: recpol_3_lowen}. The differential cross section, displayed in Fig.~\ref{fig: diffcs_3_lowen}, is quite well reproduced. As explained in Sec.~\ref{subsubsec:K0S+}, the good performance of the background models in this channel hinges entirely on the inclusion of the $K^{\ast}(1410)$ trajectory. Specifically, this trajectory provides the necessary destructive interference to counteract the sharp rise of the cross section brought about by the $K^{\ast}(892)$ contribution (Fig.~\ref{fig: diff_iso3_iso2pred}). At the most forward kaon angles, the computed cross sections can be seen to exhibit a brief increase around 2.5 GeV, not observed in the data, before dipping back down to meet their high-energy values (Fig.~\ref{fig: diff_highen_iso3_pred}). 

In Fig.~\ref{fig: recpol_3_lowen}, the $p(\gamma,K^0)\Sigma^+$ recoil
asymmetry is presented. Because only four data points are available
for $\cos\theta_K^{\ast} > 0$, this result may be considered more as a
prediction than as the actual outcome of a fit. The background
contribution equals zero because of the constant phase assumed for the
$K^{\ast}(892)$ and $K^{\ast}(1410)$ trajectories, which are the only
ingredients of the Regge model in the $K^0\Sigma^+$ channel. Real
propagators for $K^{\ast}(892)$ and $K^{\ast}(1410)$ result in a real
amplitude, and since $P$ is related to the amplitude's imaginary part,
it vanishes for this background choice. The RPR-3 model provides a slightly better overall description of $P$.

Summarizing,  Figs.~\ref{fig: diffcs_2_lowen}-\ref{fig:
  recpol_3_lowen} demonstrate the ability of the proposed RPR strategy
to provide a consistent description of the $p(\gamma,K^+)\Sigma^0$ and
$p(\gamma,K^0)\Sigma^+$ processes over an extensive energy
region. Contrary to the high-energy data, which do not unambiguously
constrain the phase of all contributing trajectories, the
resonance-region data clearly prefer a rotating phase for the $K(494)$
trajectory and a constant phase for the $K^{\ast}(892)$
one. Incidentally, the same phase options were also identified as the
most likely one in our treatment of $K\Lambda$ photoproduction~\cite{CorthalsL}. The $K^{\ast}(1410)$ phase should match the choice made for $K^{\ast}(892)$. Apart from the standard $N^{\ast}$ ``core'' set, we identified $P_{13}(1900)$ as a dominant resonant contribution. The spin-3/2 resonance $D_{33}(1700)$ was also shown to be important, as well as the spin-1/2 states $S_{31}(1900)$ and $P_{31}(1910)$, the inclusion of either of which was found to be sufficient. 

\begin{figure*}[ht]
\begin{center}
\includegraphics[width=0.7\textwidth]{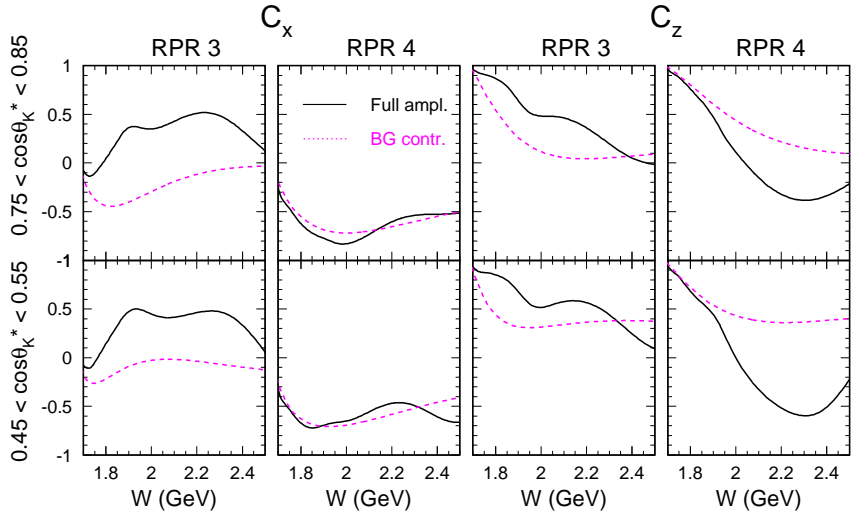}
\caption{(Color online) Energy dependence of the $p(\gamma,K^+)\Sigma^0$ double-polarization observables $C_x$ (left panels) and $C_z$ (right panels) for two bins in $\cos\theta_{K}^{\ast}$. Line conventions are as in Fig.~\ref{fig: diffcs_2_lowen}. \label{fig: doubpol_2_lowen} }
\end{center}
\end{figure*}

Finally, anticipating the publication of the $p(\gamma,K^+)Y$ double-polarization data from CLAS~\cite{Schumacher06}, we show predictions for $C_x$ and $C_z$ for the case of the $K^+\Sigma^0$ final state (Fig.~\ref{fig: doubpol_2_lowen}). It is clear that these observables are very sensitive to the choice of background model, and that the background and full RPR amplitudes lead to quite different results. 

\section{Conclusions}
\label{sec: conclusion}

We have applied the Regge-plus-resonance (RPR) strategy, developed in Ref.~\cite{CorthalsL} for $K^+\Lambda$ photoproduction, to obtain a common description of the $\gamma p \rightarrow K^+\Sigma^0$ and  $\gamma p \rightarrow K^0\Sigma^+$ processes. The framework employed involves the superposition of a limited number of $s$-channel resonances onto a reggeized $t$-channel background, resulting in a ``hybrid'' reaction amplitude valid for photon lab energies from threshold up to 16 GeV. 

For the $p(\gamma,K^+)\Sigma^0$ background amplitude, $K^+$ and
$K^{\ast +}(892)$ Regge-trajectory exchanges were assumed. We
addressed the question of whether a constant or a rotating trajectory
phase represents the optimum choice. The amount of high-energy data
proved too limited to constrain adequately the model parameters. We
therefore adopted the strategy of retaining only those Regge model variants reproducing the sign of the recoil asymmetry in the resonance region. By imposing this requirement, the option of a constant $K^+$ trajectory phase could be ruled out. 

In principle, isospin considerations allow one to transform the $K^+\Sigma^0$ photoproduction amplitude into an amplitude applicable to the $\gamma p \rightarrow K^0\Sigma^+$ process. Because the $K^0$-exchange diagram vanishes in photoproduction, the $K^0\Sigma^+$ amplitude constructed in this manner would solely comprise $K^{\ast 0}(892)$ exchange. Such a single-trajectory model, however, overshoots the $p(\gamma,K^+)\Sigma^0$ data in the resonance region. 
With one of the leading $t$-channel contributions missing, higher-mass kaon trajectories are expected to start playing a role. Including the exchange of the $K^{\ast}(1410)$ trajectory was found to lead to very good results. Due to the lack of high-energy $p(\gamma,K^0)\Sigma^+$ data, the $K^{\ast +}(1410)$ parameters and trajectory phase had to be determined against the resonance-region data. The background model constructed in this manner was extended to the high-energy domain, resulting in a prediction for $K^0\Sigma^+$ photoproduction at $\omega_{lab} = 5-16$ GeV. 

We added $s$-channel diagrams to the reggeized background amplitude. In order to minimize any double-counting effects that might arise, the number of resonances was deliberately constrained. Apart from the usual $N^*$ states $S_{11}(1650)$, $P_{11}(1710)$ and $P_{13}(1720)$ nucleon resonances, we investigated possible contributions of the two-star $P_{13}(1900)$, as well as of the $\Delta^{\ast}$ states $S_{31}(1900)$, $P_{31}(1910)$, $D_{33}(1700)$ and $P_{33}(1920)$. 

Remarkably, the background option assuming rotating $K$ and constant $K^{\ast}$ trajectory phases, along with a negative $K^{\ast}(892)$ tensor coupling, was found to consistently provide the best fit to the data, whichever resonance contributions were added. The alternative parameterization with rotating phases for all trajectories, which has always been regarded as the ``standard'' choice for the high-energy description, turned out to be incompatible with the resonance-region data. 
Apart from the standard $N^{\ast}$ ``core'' states $S_{11}(1650)$, $P_{11}(1710)$ and $P_{13}(1720)$, the two-star $P_{13}(1900)$ was identified as a necessary contribution, irrespective of background assumptions. For the $\Delta^{\ast}$s, the situation is less clear. When assuming the ``preferred'' background model described above, including either of the spin-1/2 resonances $S_{31}(1900)$ or $P_{31}(1910)$ turned out to be sufficient, whereas the spin-3/2 $D_{33}(1700)$ state was found to be considerably more important than $P_{33}(1920)$. These conclusions are, however, closely linked to the chosen background parameterization.  

    Above all, it should be realized that pinpointing the dominant
  $s$-channel diagrams as yet remains a delicate business. While the
   inclusion of extra resonances invariably leads to a decrease in
 $\chi^2$, this does not automatically imply an increased likelihood
 for the constructed model. Furthermore, it is seldom clear whether a
 similar agreement with the data cannot be obtained using a different
   combination of states. Indeed, as the number of model parameters
   increases, it becomes ever harder to check whether the attained
  minimum in $\chi^2$ is truly a global minimum, and whether or not
other such minima exist. Evidently, this challenge will prove even
   more daunting in coupled-channels models than at tree level. It
 will, however, have to be addressed carefully in future analyses of
weak channels such as kaon photoproduction.

\begin{acknowledgments}
We wish to thank M. Guidal for providing very helpful feedback. We also acknowledge enlightening discussions with M. Vanderhaeghen. This work was supported by the Fund for Scientific Research, Flanders (FWO) and the research council of Ghent University.
\end{acknowledgments}

\appendix* 

\section{Isospin symmetry in the $\bm{K^+\Sigma^0}$/$\bm{K^0\Sigma^+}$ channels}
\label{app: isospin}
In this Appendix, we sketch how isospin arguments can be applied to establish relations between the coupling constants for the $\gamma p \rightarrow K^+ \Sigma^0$ and $\gamma p \rightarrow K^0 \Sigma^+$ channels. Only the relations specifically required for this work are mentioned. A more extensive review can e.g.~be found in Ref.~\cite{Stijn_sigma}. In what follows, the isospin symmetry of the various meson and baryon multiplets is assumed to be exact.

All hadronic decay processes relevant to the RPR treatment of forward-angle $K \Sigma$ photoproduction are either of the type $N \rightarrow K \Sigma$ or $\Delta \rightarrow K \Sigma$. 
Because of the isovector nature of the $\Sigma$ particle, the hadronic couplings are proportional to the Clebsch-Gordan coefficients:
\begin{align}
g_{K\Sigma N} \quad &\sim \quad <I_K=\frac{1}{2},~M_K ;~I_{\Sigma}=1,~M_{\Sigma}~ |~I_{N} = \frac{1}{2},~M_N >\,, \\
g_{K\Sigma \Delta} \quad &\sim \quad <I_K=\frac{1}{2},~M_K ;~I_{\Sigma}=1,~M_{\Sigma}~ |~I_{\Delta} = \frac{3}{2},~M_{\Delta} >\,.
\end{align}
When adopting the following conventions for the isospin states of the $N$, $K$ and $\Sigma$ particles,
\begin{align}
& \Sigma ^+  \; \leftrightarrow \;  - \left|I=1, \; M =+1 \right> \;,\nonumber \\
&\Sigma ^0  \; \leftrightarrow \;  + \left|I=1, \; M = 0 \right> \;, \nonumber\\
&\Sigma ^-  \; \leftrightarrow \;  + \left|I=1, \; M =-1 \right> \;;  \\ 
&p \leftrightarrow \left|I=\frac{1}{2}, \; M =+\frac{1}{2} \right> \leftrightarrow K^+\;,\nonumber \\
&n \leftrightarrow \left|I=\frac{1}{2}, \; M =-\frac{1}{2} \right> \leftrightarrow K^0\;.\nonumber \\
\end{align}
these simple relations emerge:
\begin{align}
g_{K^0 \Sigma^+ p}\ &= \sqrt{2} \ g_{K^+ \Sigma^0 p} \;, \\
g_{K^0 \Sigma^+ \Delta^+}\ &= -\frac{1}{\sqrt{2}} \ g_{K^+ \Sigma^0 \Delta^+}\,.
\label{eq:SKN}  
\end{align}

Contrary to the hadronic parameters, the relations between electromagnetic couplings have to be distilled from experimental information. In principle, the value of the magnetic transition moment $\kappa_{K^* K}$ can be determined on the basis of the proportionality $\kappa_{K^* K}^2 \sim \Gamma_{K^* \rightarrow K \gamma}$. Within the context of tree-level models, however, the coupling constants are frequently considered as ``effective couplings'' in which, for example, part of the final-state interaction effects are absorbed.  It is a common procedure to use only the ratios of the measured decay widths to connect isospin-related coupling constants. Using the PDG values for the $K^{*+}(892)$ and $K^{*0}(892)$ widths, i.e.~\cite{PDG04}:
\begin{eqnarray}
\Gamma_{K^{*+}(892) \rightarrow K^+ \gamma} &=& 50 \pm 5 \mbox{~keV} \;, \label{eq:gammaK+} \\
\Gamma_{K^{*0}(892) \rightarrow K^0 \gamma} &=& 116 \pm 10 \mbox{~keV} \;, \label{eq:gammaK0}
\end{eqnarray}
the following expression is obtained:
\begin{equation}
\kappa_{K^{*0}(892)\, K^0} = - 1.52\  \kappa_{K^{*+}(892)\, K^+} \;.
\label{eq:K*+K*0}
\end{equation} 
The relative sign in the last expression was selected on the basis of a constituent-quark model prediction by Singer and Miller~\cite{Singer}, which accurately reproduces the experimental widths of Eqs.~(\ref{eq:gammaK+}) and~(\ref{eq:gammaK0}).


\begin{thebibliography}{99}
\bibitem{corr_lagr} We wish to rectify an inaccuracy in our previous work on $K\Lambda$ photoproduction. The Lagrangians for coupling to spin-3/2 resonances quoted in Ref.~\cite{CorthalsL} correspond to the standard normalization, which is also assumed in this work. However, the spin-3/2 coupling constants quoted in Table IV of Ref.~\cite{CorthalsL} were in fact obtained with a different normalization, namely:
\begin{equation}
\begin{split}
\mathcal{L}_{\gamma p N^{\ast}(\frac{3}{2})} &=~i \ \frac {e\kappa_{\ssst pN^{\ast}}^{\left( 1
\right) }}{M_{N^{\ast}}+Mp}\ 
\overline{R}^\mu \theta_{\mu \nu} \left(Y \right) 
\Gamma_\lambda N F^{\lambda \nu} \\ 
&- \frac{e \kappa_{\ssst pN^{\ast}}^{\left( 2
\right)} }{(M_{N^{\ast}}+Mp)^2}\ \overline{R}^\mu \theta_{\mu \nu} \left( X \right)
\Gamma \left( \partial_\lambda N \right) F^{\nu \lambda} + h.c.,
\end{split}
\end{equation}
\begin{equation}
\mathcal{L}_{K \Lambda N^{\ast}(\frac{3}{2})} = \frac{f_{\ssst K \Lambda N^{\ast}}}{M_{\ssst N^{\ast}}}
\overline{R}^\mu \theta_{\mu\nu} \left( Z \right) \Gamma' \Lambda \left 
( \partial^\nu K \right) + h.c.\,,
\label{eq:gpr_3_2_old}
\end{equation}
with the PDG values used for $M_{N^{\ast}}$. While this does not affect any of the conclusions drawn in Ref.~\cite{CorthalsL}, it should be kept in mind when comparing the values of the reported coupling constants to those obtained with other models.
\end{thebibliography}
\end{document}